\documentclass[jgrga,arxiv]{agutex4}
\usepackage{amsmath, amssymb}
\usepackage{color}
\usepackage{graphicx}
\usepackage{lineno}

\setkeys{Gin}{draft=false}
\pdfoutput=1
\authorrunninghead{HANGX AND BRANTUT}

\titlerunninghead{MICROMECHANICS OF AGGREGATE COMPACTION}

\authoraddr{Corresponding author: S.J.T. Hangx, High Pressure and Temperature Laboratory, Department of Earth Sciences, Faculty of Geosciences, Utrecht University, Budapestlaan 4, 3584 CD, Utrecht, the Netherlands. (s.j.t.hangx@uu.nl)}

\begin{document}

\title{Micromechanics of high pressure compaction in granular quartz aggregates}

\authors{Suzanne J.T. Hangx,\altaffilmark{1} Nicolas Brantut\altaffilmark{2}}

\altaffiltext{1}{Department of Earth Sciences, Faculty of Geosciences, Utrecht University, Utrecht, the Netherlands.}

\altaffiltext{2}{Department of Earth Sciences, University College London, London, UK.}

\begin{abstract}
  The mechanical behaviour of porous sandstones is generally modelled using concepts from granular mechanics, often overlooking the effect of cementation. To probe the key differences between sand and sandstone mechanics, we performed triaxial deformation experiments on Ottawa quartz sand at $5-40$ ~MPa effective confining pressure. At $5$ ~MPa, the samples are able to dilate. At higher confinement, the aggregates show continuous compaction, displaying strain hardening. The stress-strain behaviour is non-linear and the exact onset of inelastic compaction could not be determined accurately. Measured P-wave velocities show the development of anisotropy. With increasing axial strain, the along-axis velocities tend to increase, while velocities perpendicular to the compression axis tend to decrease (at low pressure) or remain constant (at high pressure). In samples deformed under elevated pressure conditions, acoustic emission event locations are diffuse. Microstructural investigations show an increase in grain chipping and crushing with increasing confining pressure, but no evidence of localised compaction could be observed. The nature of the pore-fluid, either decane or water, does not significantly influence the mechanical behaviour at strain rates of $10^{-6}$ to $10^{-4}$ $s^{-1}$. Grain angularity and grain-size distribution also did not significantly change the mechanical behaviour. We infer that our observations indicate that the lack of cementation introduces additional degrees of freedom for grains to slide, rotate, and reorganise at the sample-scale, precluding the existence and sustainability of stress concentrations beyond the grain-scale. This results in progressive compaction and hardening, and lack of compaction localisation.
\end{abstract}

%
%

\begin{article}

%
%

\section{Introduction}
Highly porous unconsolidated sands and poorly consolidated sandstones (porosity $\geq 20\%$) constitute important reservoirs targeted for hydrocarbon extraction (e.g. Gulf of Mexico \citep{Ostermeier1995, Li2003}; Slochteren Sandstone, Groningen Gas Field, the Netherlands \citep{Hettema2002}). It is their high porosity which makes such reservoirs suitable for fluid production due to their often high permeability. However, changes in effective overburden stress due to fluid extraction can result in compaction at the reservoir level. Even relatively small amounts of compactive strain (few tenths of a percent) can lead to surface subsidence and induced seismicity \citep{Doornhof2006, Spetzler2017, Pijnenburg2018, Pijnenburg2019}, while larger strains (several percent) and strain localisation can lead to geotechnical problems, such a wellbore casing collapse \citep{Doornhof2006, Li2003}, or reduce reservoir permeability impacting production \citep{Olsson2002}. Compaction is most likely caused by a direct poro-elastic response \citep{Wang2000} and time- or rate-independent inelastic compaction, as well as time-dependent compaction creep, which may continue long after production has ceased \citep{Doornhof2006}.  

Porous sandstones have been extensively studied and a vast body of work is dedicated to deformation and strain localisation in sandstone \citep[see][and references therein]{Wong2012}. At very low confining pressure, strain localisation in porous sandstones has been shown to occur via dilatant shear localisation moving to compactant shear localisation at low to moderate confinement \citep{Wong2012}. At higher confining pressures, compaction may transition towards strain localisation via discrete \citep{Fortin2007, Tembe2008, Louis2006, Heap2015, Klein2001} or diffuse compaction bands \citep{Baud2004, Olsson2000, Olsson2002} to distributed cataclastic flow under hydrostatic conditions \citep{Baud2004}. While discrete compaction bands (i.e. $\leq$~3 grains wide \citep{Baud2004}) may be formed under both constant strain rate and constant stress conditions \citep{Heap2015}, they have been observed in very few sandstones deformed under laboratory conditions. It is believed that discrete compaction bands are promoted in high porosity ($\geq 20\%$), well-sorted and spatially homogeneous sandstones \citep{Cheung2012}, such as the Bentheim, Bleurswiller and  Diemelstadt sandstone \citep{Fortin2007, Tembe2008, Louis2006, Heap2015}. Furthermore, under wet conditions sandstones generally tend be significantly weaker \citep{Baud2000, Duda2012}. 

Studies on the uniaxial consolidation (porosity loss) behaviour of sands are numerous, focusing on the effect of grain-size and angularity, mineralogy and chemical environment \citep{Brzeso2014, Chuhan2003, Chuhan2002, Mesri2009, Fawad2011, BrzesoBrantut2014}. These studies show that higher initial porosity, increasing grain angularity, increasing grain-size and size uniformity, decreasing intrinsic grain strength and the presence of aqueous fluids enhance consolidation \citep{BrzesoBrantut2014, Brzeso2014, Hangx2010, Chuhan2003, Chuhan2002, Lee1967, Vesic1968, Zoback1976}. Initial consolidation of sand is attributed to tighter packing promoting particle locking, while upon higher stresses interparticle slip and particle crushing and fracturing lead to further particle unlocking and concomitant compaction \citep{Mesri2009, Munoz-Ibanez2018}. Additionally, many studies exist in the soil mechanics community, regarding triaxial consolidation of sands and clays, though often at low confining pressures \citep[e.g. see][]{Pestana2002, Alikarami2015}. Under low confining conditions (confining pressures of $0.1-7$~MPa), recent experiments using X-ray computed tomography show the formation of dilation and compaction shear bands \citep{Alikarami2015}, accommodated by grain rolling and frictional sliding \citep{Alikarami2015, Karner2005}. Grain angularity impacts compaction and strain localisation behaviour by enhancing interlocking in more angular aggregates, which in turn inhibits dilation \citep{Alikarami2015, Guo2007}. However, systematic experimental study of the micromechanics of strain localisation in unconsolidated sands at higher confinement has received much less attention \citep{Karner2005, Skurtveit2013, Nguyen2014}. 

Instead, extensive numerical modelling efforts have been made to predict the behaviour of compacting sands and sandstones, either through constitutive modelling \citep{Issen2000, Pestana1999, Tengattini2014, Choo2018, Buscarnera2014, Einav2007} or employing particle models such as Discrete Elements \citep{Potyondy2004, Wang2008, Wu2018, Marketos2009}. Such numerical models are crucial in providing the means to extrapolate granular material behaviour across spatial and temporal scales. However, they generally assume that rocks can be approximated as grains held together by cement bonds \citep[e.g. see][]{Einav2007, Potyondy2004}. This would imply that once sufficient cement bonds are broken, the material would essentially behave as a loose granular aggregate, as confirmed by low confinement experiments on loose sand and artificially cemented sand \citep{Bernabe1992}. While compaction bands in sand have not been experimentally verified, DE modelling efforts have suggested that it is possible to generate compaction bands in sand packs, though less pronounced than in sandstones \citep{Marketos2009}.

Given the first-order similarities in mechanical behaviour between sands and sandstones, and the effectiveness of granular microphysical models to explain the deformation of sandstones, should imply that unconsolidated sand aggregates behave in the same way as sandstones. This would mean that in sand the onset of inelastic compaction is driven by the onset of grain crushing at the microscale. It also implies that shear bands will develop at low confinement \citep{Wong2012}, while compaction bands should be observed at high confining pressure ($P_\mathrm{c}$) if the grain-size is narrowly distributed \citep{Wong2012, Cheung2012}. It should be noted that though this implies the assumption of a double-yield cap model, it has been shown that upon deformation of porous aggregates, and the concomitant compaction and porosity reduction, the initial yield cap can expand \citep[e.g.][]{Wong1992, Xiao2003, Bedford2018, Pijnenburg2019}. Essentially this means that any material has the potential to transition from one side of the yield cap to the other, while the yield cap expands, regardless of how the material is deforming. Experiments have already shown that at low confinement, dilatant shear bands do form in loose sand \citep{Alikarami2015}, similar to porous sandstones \citep{Wong2012}. However, while sand experiments performed at higher confining pressure show significant grain crushing, as also seen seen in sandstones \citep{Karner2005, Zoback1976}, the micromechanical characteristics of strain localisation in this regime remain unclear.

We performed a series of triaxial deformation experiments on homogeneous, highly porous, unconsolidated quartz sand aggregates at effective confining pressures in the range $5--100$~MPa. During the experiments, the effect of chemical environment (decane versus water), grain angularity (well-rounded versus subrounded/-angular) and grain-size distribution (narrow versus broad) on deformation behaviour is studied. We aim to highlight systematic qualitative and quantitative differences between our results on sand and observations on cemented sandstones, by making a systematic comparison between high-pressure deformation of sand versus sandstone, with a special focus on microstructural controls, strain localisation and acoustic characteristics.

%
%

\section{Experimental methods}
We performed two types of conventional triaxial experiments on fluid-saturated sand aggregates under room temperature conditions: 
\begin{enumerate}
    \item Two hydrostatic tests (OS--04 and OS--17) were done at $P_\mathrm{c}$ up to $55$~MPa and $105$~MPa, respectively. Decane-saturated test OS--04 was performed at fixed confining pressure steps of $2.5$~MPa, while during each step the sample was allowed to reach a near-steady volumetric strain rate of \textless $10 ^{-6}$~s$^{-1}$. Water-saturated test OS--17 was performed at a fixed confining pressure step rate of $2.5$~MPa/$10$~mins. Both experiments were performed on a grain-size batch of $350 \pm 50 \mu$m.
    \item Sixteen triaxial deformation tests were done with applied confining pressures ($P_\mathrm{c}$) in the range $10-45$ MPa, at a constant axial strain rate ($\dot{\epsilon}$) of $10^{-5}$~s$^{-1}$. After $\sim 5\%$ axial strain, strain rate was cycled several times from 10$^{-6}$~s$^{-1}$ to 10$^{-4}$~s$^{-1}$, to test the effect of strain rate on deformation. These tests were performed using three different grain-size batches ($350 \pm 50$~$\mu$m, $340 \pm 160$~$\mu$m and $240 \pm 60$~$\mu$m).
    
\end{enumerate}

The experiments were either performed using decane or water as pore-fluid (pore-fluid pressure, $P_\mathrm{p}$ = $5$~MPa), to test the effect of chemical environment on deformation behaviour. A summary of the experiments, plus key parameters obtained during the tests, is presented in Tables \ref{Tab:DecWat} and \ref{Tab:GSD-shape}.

\begin{table}
    \caption{Conventional triaxial compression and hydrostatic experiments performed on fluid-saturated Ottawa sand aggregates ($d = 300-400 \mu$m) at $20 ^\circ$C and $P_\mathrm{p}$ = $5$~MPa, plus unloading Young's Modulus obtained in the tests. Additionally, mean grain size values are given when measured.}
    \label{Tab:DecWat}
    \centering
    \begin{tabular}{l c c c c c}
    \hline
 Sample                         & $\phi_\mathrm{0}$      & $\phi_\mathrm{i}$\tablenotemark{a}   & $P_\mathrm{c}^\mathrm{eff}$   & $E_\mathrm{u}$  & $d_{50}$\\
                                & [\%]          & [\%]          & [MPa]          & [GPa]          & [$\mu$m]\\
\hline
\hline
starting material &   &   &   &   & 427-444 \\
$  $ \\
\hline
\multicolumn{5}{l}{Decane-saturated sand aggregates} \\
\hline
  
  OS--03                        & 36.3          & 33.6          & 20            & 6.3             & 351-375            \\
  OS--05                        & 36.2          & 33.5          & 25            & 6.8             & 351-353            \\
  OS--02                        & 36.0          & 32.6          & 30            & 7.8             & 318-339            \\
  OS--04\tablenotemark{b}       & 36.1          & 36.1\tablenotemark{c}         &5--50 &    & 410-427\\
  $ $ \\
\hline
\multicolumn{5}{l}{Water-saturated sand aggregates} \\
\hline
  OS--15                        & 36.3          & 36.3\tablenotemark{c} & 5     & 3.6             &             \\
  OS--16\tablenotemark{d}       & 36.2          &         & 10            & 4.5             & 438-439            \\
  OS--08                        & 36.1          & 34.7          & 15            &   & 364-374\\
  OS--11\tablenotemark{e}       & 36.3          & 34.3          & 20            & 5.7             &             \\
  OS--09\tablenotemark{f}       & 36.1          & 34.0          & 20            & 5.4             &              \\
  OS--06                        & 34.6          & 32.9          & 20            & 5.9             &             \\
  OS--12                        & 36.0          & 33.9          & 20            & 6.5             & 374-386 \\
  OS--07                        & 36.3          & 33.8          & 25            & 6.7             & \\
  OS--10                        & 36.2          & 32.7          & 30            & 7.3             & 300 \\
  OS--18                        & 36.3          & 31.1          & 40            & 8.1             &             \\
  OS--17\tablenotemark{b}       & 36.3          & 36.3\tablenotemark{c} & 5--100    &   & \\
  $ $ \\
\hline
\end{tabular}
\tablenotetext{}{Symbols: $\phi_\mathrm{0}$ is sample porosity prior to set-up, $\phi_\mathrm{i}$ is the estimated sample porosity after application of $P_\mathrm{c}$ and $P_\mathrm{p}$, $P_\mathrm{c}^\mathrm{eff}$ represents effective confining pressure, $E_\mathrm{u}$ is the Young's modulus obtained from a linear fit to the linear (elastic) part of the stress--strain curve during unloading, $d_{50}$ is the mean grain size obtained from particle size analysis (two measurements).}
\tablenotetext{a}{The estimated porosity after application of confinement does not include any sample volume changes during the pore pressure application phase. Error is  $\sim 0.6\%$ (see Appendix \ref{AppendixA}).}
\tablenotetext{b}{Hydrostatic experiments: OS--04 performed at $P_\mathrm{c}$ steps of $2.5$~MPa, while during each step the sample was allowed to reach a near-steady volumetric strain rate of $\textless$ $10^{-6}$~s$^{-1}$. OS--17 performed at a fixed confining pressure step rate of $2.5$~MPa/$10$~mins.}
\tablenotetext{c}{Experiment started immediately after application of the pore pressure ($P_\mathrm{c}^\mathrm{eff}$ = $5$~MPa).}
\tablenotetext{d}{Sample experienced jacket leakage, so no pore volume changes could be obtained.}
\tablenotetext{e}{Sample deformed up to 1.5$\%$ $e$.}
\tablenotetext{f}{Sample deformed up to 4$\%$ $e$.}
\end{table}

\subsection{Sample material and preparation}
The sand used in $17$ of the experiments was ASTM standard C778 Ottawa sand obtained from US Silica (Ottawa, IL, USA). Specifications indicated a SiO$_2$ content of 99.7 wt-$\%$, with minor quantities of Al$_2$O$_3$ ($0.06$~wt-$\%$), Fe$_2$O$_3$ ($0.02$~wt-$\%$) and TiO$_2$ ($0.01$~wt-$\%$). Scanning electron microscopy (SEM) showed that most sand grains are well-rounded, with smooth surfaces characterised by pressure solution indentations and quartz overgrowths. Fractions of the sand ($350 \pm 50$~$\mu$m, $340 \pm 160$~$\mu$m and $240 \pm 60$~$\mu$m) were prepared by sieving twice the material as received. Additionally, one experiment was performed using Beaujean sand, the same as used by \citet{Brzeso2011, Brzeso2014, BrzesoBrantut2014}. This material was obtained from the Heksenberg Formation, Beaujean quarry, Heerlen, Netherlands, and sieved to a grain-size fraction of $240 \pm 60$~$\mu$m. Compared to Ottawa sand, the Beaujean sand is more subrounded to subangular.

\begin{figure*}
    \centering
    \includegraphics[scale=0.8]{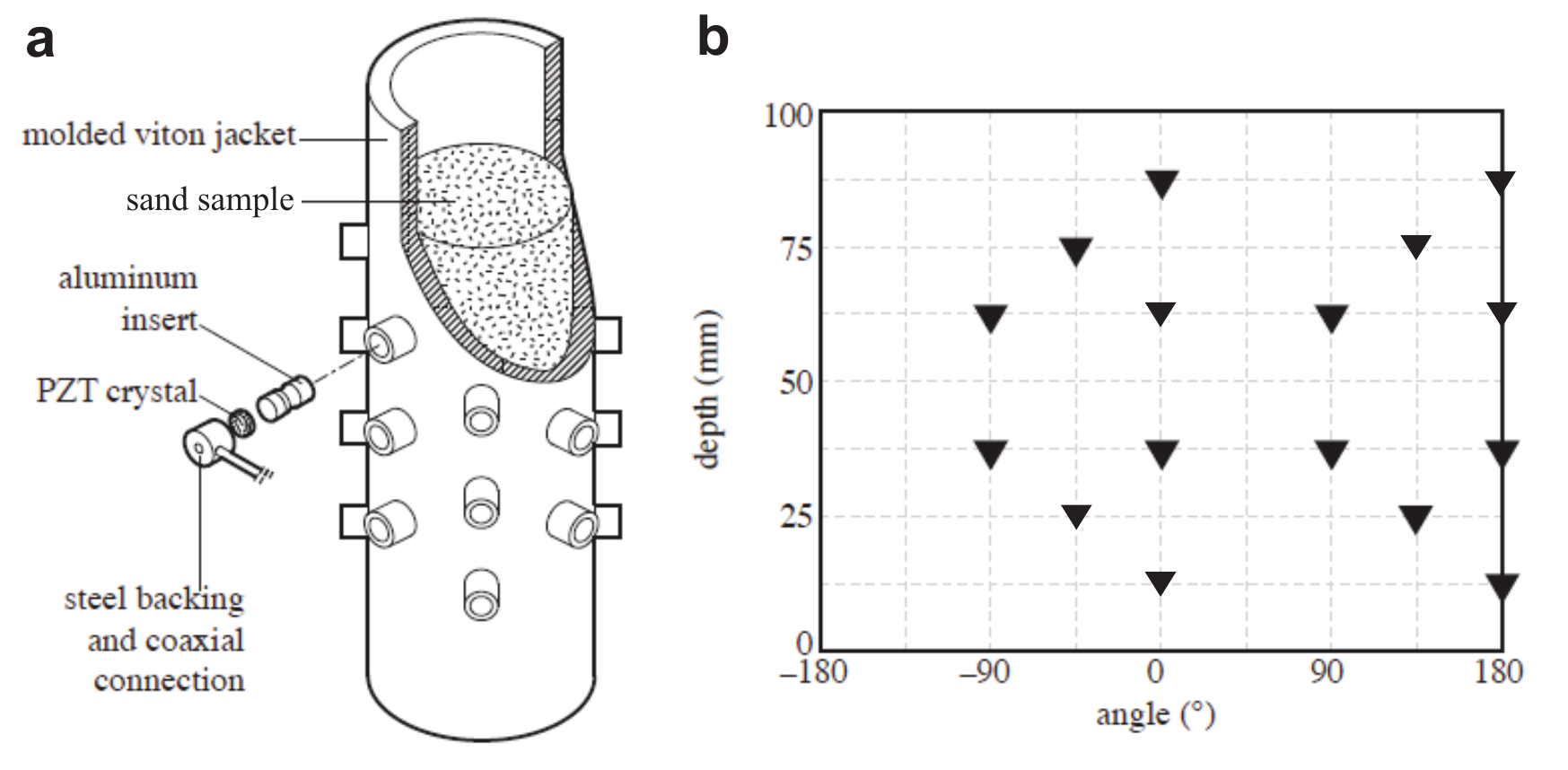}
    \caption{a) Schematic of the sample (diameter is 40 mm and length is 100 mm) and sample jacket. In total 16 acoustic transducers were mounted on molded inserts in the jacket. Unused inserts were blanked with aluminium plugs. b) Map of sensor positions around the sample.}
    \label{Fig:Set-up}
\end{figure*}

Pre-packed sand aggregates ($40$~mm diameter, $100$~mm length) were prepared by pouring a fixed amount of sand ($\sim 250$~g in mass) into a Viton sample jacket used for the experiments (see Figure \ref{Fig:Set-up}), followed by $10$~mins on a shaking plate with a small weight applied to the sand pack. Starting porosity was calculated using the exact sample dimensions and sample mass, assuming a density of quartz of $2.66$~g/cm$^3$. This preparation method resulted in a well-controlled starting porosity within the range $36.0-36.3 \%$ for the Ottawa grain-size fraction of $300-400$~$\mu$m (Table \ref{Tab:DecWat}). Note that sample OS--06 had a slightly lower initial porosity of $34.6$\%, which could affect its mechanical behaviour \citet{Brzeso2014}.

\subsection{Experimental apparatus and procedure}
The deformation experiments were performed in a conventional triaxial apparatus at the Rock and Ice Physics Laboratory of University College London \citep[see description in][]{Eccles2005, Brantut2014}. The machine consists of an oil-filled pressure vessel, with servo-controlled application of axial load and pore pressure, and manual application of confining pressure. After filling of the vessel, confining pressure on the sample was $\sim 5$~MPa. Confining and pore pressures were then increased in $1$~MPa steps, such that the pore pressure never exceeded the confining pressure, up to $10$~MPa confining pressure and $5$~MPa pore pressure. Subsequently only confining pressure was increased in $5$~MPa steps to the target conditions. The sample was then left to equilibrate for about $2$~hours. Axial shortening was measured using the average of two linear variable differential transformers (LVDTs), located outside of the pressure vessel, and used to calculate axial strain. All tests were performed under drained, constant pore pressure conditions. After $\sim 5$~\% axial strain, strain rate-stepping stages were performed by down-stepping the strain rate by one order of magnitude ($\dot{\epsilon}$ = $10^{-6}$~s$^{-1}$), followed by up-stepping the strain rate by one order of magnitude compared to the initial strain rate ($\dot{\epsilon}$ = $10^{-4}$~s$^{-1}$). This sequence was repeated. 

For all experiments, total axial strains of $\sim 7.5~\%$ were achieved. Pore volume changes were determined by measuring the change in pore-fluid volume during deformation, as determined from the position of the actuator inside the servo-controlled pore pressure intensifier. The triaxial deformation experiments were terminated by halting the loading ram. Subsequently, the sample was unloaded at a strain rate of 10$^{-5}$~s$^{-1}$, followed by reducing the confining and pore pressures to room pressure, making sure that at all times $P_\mathrm{c}$ \textgreater $P_\mathrm{p}$.

\begin{table*}
\caption{Conventional triaxial compression experiments performed on fluid-saturated aggregates of Ottawa and Beaujean sand, with different grain angularities and grain-size distributions (GSD), at 20$^\circ$C and $P_\mathrm{p}$ = 5 MPa, plus unloading Young's Modulus obtained in the tests. Additionally, mean grain size values are given when measured.}
\label{Tab:GSD-shape}
\centering
\begin{tabular}{l c c c c c c}
\hline
 Sample                         & Sand type         & $d$      & $\phi_\mathrm{0}$        & $\phi_\mathrm{i}$          & $E_\mathrm{u}$    &$d_{50}$\\
                                &                   & [$\mu$m]          & [\%]              & [\%]                & [GPa]    & [$\mu$m]  \\
\hline
\hline
\multicolumn{6}{l}{Effect of grain angularity: decane-saturated sand aggregates} \\
\hline
  OS starting material     &   &   &   &   & 308-309 \\
  BS starting material     &   &   &   &   & 294-298 \\
  OS--19                        & Ottawa            & 180--300           & 37.4             & 30.9                & 8.0       & 280-283\\
                                & (well-rounded) \\                        
   
  BS--01                        & Beaujean          & 180--300           & 38.6             & 31.6                & 8.0       &198-227\\
                                & (subrounded/-angular) &                         \\
  $ $ \\
\hline
\multicolumn{6}{l}{Effect of grain-size distribution: water-saturated sand aggregates} \\
\hline
  starting material (narrow)     &   &   &   &   &  & 308-309 \\
  starting material (broad)     &   &   &   &   &   & 403-450 \\
  OS--07                        & Ottawa            & 300--400            & 36.3             & 33.8               & 6.7    & \\
                                 &                  & (narrow GSD) \\
  OS--14                        & Ottawa            & 180--500            & 34.9             & 32.5               & 5.5     & 308-335\\
                                 &                  & (broad GSD) \\
  $ $ \\
\hline
\end{tabular}
\tablenotetext{}{Symbols: $\textit{d}$ is grain-size, $\phi_\mathrm{0}$ is sample porosity prior to set-up, $\phi_\mathrm{i}$ is sample porosity after application of $P_\mathrm{c}$ and $P_\mathrm{p}$, $P_\mathrm{c}^\mathrm{eff}$ represents effective confining pressure, $E_\mathrm{u}$ is the Young's modulus obtained from a linear fit to the linear (elastic) part of the stress--strain curve during unloading, $d_{50}$ is the mean grain size obtained from particle size analysis (two measurements).}
\end{table*}

\subsection{Wave velocity and acoustic emission measurements}
During deformation, acoustic emissions (AEs) and the evolution of P-wave velocities with increasing deformation were monitored. The sample jacket is equipped with 16 piezoelectric transducers positioned around, and directly touching, the sample, and connected to 40~dB high-frequency preamplifiers and a 50~MHz digital recording system \citep[e.g.,][]{Brantut2014} (Figure \ref{Fig:Set-up}). P-wave velocities were determined by regularly and sequentially sending a high frequency ($1$~MHz) and high voltage ($250$~V) pulse on each transducer while recording transmitted waves on the remaining sensors. During the time intervals between active wave velocity measurements, we monitored AE activity and recorded incoming signals when a threshold voltage (typically about 100 mV) was detected over five sensors. In addition to wave velocity and triggered AE waveform data, we also monitored the total of AE hits on each individual sensor, with time bins of $5$~s.

\subsection{Data processing}
External axial load, piston displacement, confining pressure, sample temperature, pore-fluid pressure and pore-fluid volume change signals were logged at time intervals of $1$~s. The raw data were processed to give differential stress ($\sigma_1 - \sigma_3$), axial strain ($\epsilon$), axial strain rate ($\dot{\epsilon}$) and porosity change ($\Delta \phi$) data versus time. The displacement data were corrected for apparatus distortion using predetermined stiffness calibrations. The change in pore volume was used to calculate porosity change, assuming rigid solid particles.

Using a cross-correlation technique, precise P-wave arrival times were determined \citep[for details, see][]{Brantut2014}. Assuming that volumetric deformation was uniformly distributed along the sample axis, the sensor positions were corrected from the axial and radial deformation achieved in each experiment. The local positions of the transducers around the sample allow for P-wave speeds to be calculated in four directions relative to the sample axis: $90.0 ^{\circ}$, $58.0 ^{\circ}$, $38.7 ^{\circ}$, and $28.1 ^{\circ}$, respectively. AEs were located using the arrival times of waveforms to the 16 transducers, assuming a homogenised transversely isotropic P-wave speed model that uses P-wave speeds interpolated in time from the measured values. Due to severe clipping of the wave form, it was not possible to assess the cumulative AE energy.

\subsection{Microstructural methods} \label{Micro-methods}
After the experiments, the jacketed samples were extracted from the pressure vessel. Due to the friable nature of the sand aggregate, intact extraction of the sample was only possible by freezing the sample at $-20 ^\circ$C for $\sim 12$ hrs, while inside the jacket. The frozen sand aggregate was subsequently removed from the sample sleeve and emplaced in a fluorinated ethylene propylene (FEP) jacket of the same dimensions as the frozen sample. The samples were allowed to defrost and dry in an oven at $80~^\circ$C for several weeks, prior to impregnation with LR White epoxy resin. Subsequently, thin sections were made from the material, roughly at the centre of the aggregate, parallel to the compression axis. Though the samples were impregnated they remained friable and during thin section preparation some grains were plucked from the section.

\begin{figure*}
    \centering
    \includegraphics[width= 10cm]{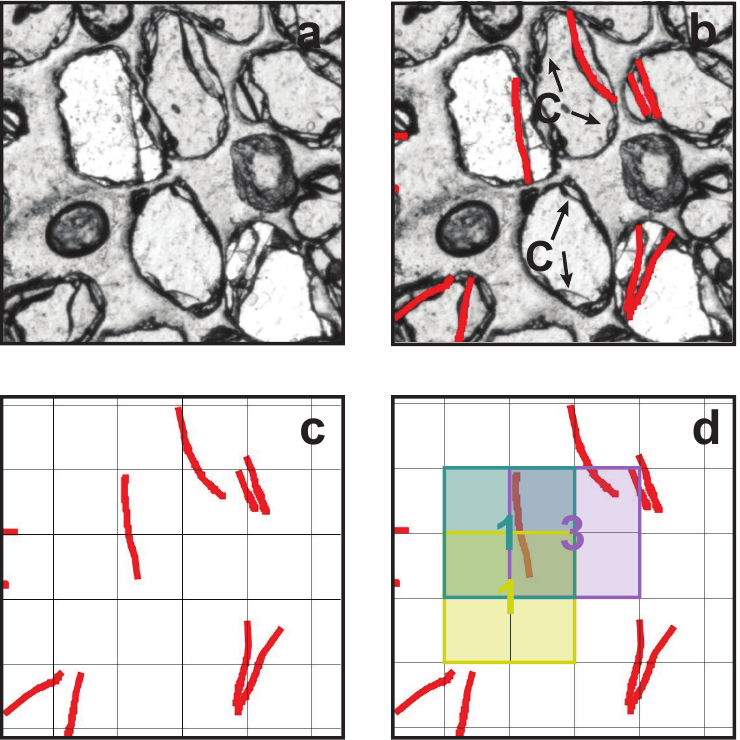}
    \caption{Procedure for processing of the microstructures to produce crack density contour maps. a) Optical micrograph of a portion of a typical sand microstructure. b) Cracks are marked in red, while chipping-related cracks (C) are not taken into account. c) The crack map is overlain by a grid of $100$~$\mu$m spacing. d) Within squares of four grid cells, the number of cracks is counted and attributed to the centre of the square. The resulting columns and rows are converted to a crack density contour map.}
    \label{Fig:Crack_map}
\end{figure*}

Using an optical microscope, micrographs of approximately $10$ by $10$ mm were made. The freezing procedure did not seem to have affected the microstructure, as many grains and their associated grain fragments appear to be undisturbed. From these micrographs, crack maps were constructed (see Figure \ref{Fig:Crack_map}) by manually tracing all through-going or partially through-going cracks visible within the sand grains. Small cracks along grain edges, caused by chipping, were not mapped. The resulting crack maps were overlain by a grid of $100$~$\mu$m spacing. Within a square of four cells, the number of cracks was counted and attributed to the centre of the square. Each new square was shifted one cell (quadrant) with respect to the previous one (analogous to a Kalsbeek counting net \citep{Kalsbeek1963}). The resulting columns and rows of numbers were converted to a crack density contour map.

Due to the low freezing temperature of decane ($-30~^\circ$C) it was not possible to freeze the decane-saturated samples, following the same method. Therefore, no attempts were made to preserve the microstructures of the decane-filled samples, and the sand was simply allowed to dry after removal from the sample sleeve. Loose sand grains from such experiments were examined using Scanning Electron Microscopy (SEM), by mounting the powder on double-sided carbon tape on stubs.

Undeformed starting material and selected, post-deformation sand samples were analysed using a Malvern laser particle sizer, to determine changes in the average grain-size ($d_\mathrm{50}$) and grain-size distribution of each sample due to deformation. Prior to sampling, each material was well mixed and a $0.5$~g sample was taken for particle analysis. Agglomeration effects can appear to lead to Malvern particle size analysis overestimating the mean grain-size by up to $\sim 1.5$ times \citep{Hangx2010}, as well as to post-experiment distributions with grain-sizes that are larger than those observed in the starting material. However, compared to grain-size analysis via optical methods, this method provides the opportunity to sample a larger number of grains, plus circumvents any stereological uncertainty obtained from 2D thin section analysis.

%
%

\section{Results}

\subsection{Fluid-saturated Ottawa sand experiments: effect of chemical environment} \label{main-OS-expts}
\subsubsection{Mechanical data} \label{Mech_data}
In this paper, we adopt the convention that compressive stresses, compressive axial strains and porosity change are measured positive. The principal compressive stresses are denoted as $\sigma_i$, with $\sigma_1 > \sigma_2 = \sigma_3 = P_\mathrm{c}$. The effective principal stresses are denoted $\sigma^\mathrm{eff}_\mathrm{i}$, and are defined as $\sigma^\mathrm{eff}_\mathrm{i}$ = $\sigma_i$ -- $P_\mathrm{p}$, where $P_\mathrm{p}$ is pore pressure. Porosity changes during each experiment are measured from the point that pore-fluid pressure is applied ($P_\mathrm{c}^\mathrm{eff}$ = $5$~MPa). Note that no porosity change can be measured for the increase in effective confining pressure to 5 MPa, as up to that point the sample is dry. Extrapolation of porosity-$P_\mathrm{c}^\mathrm{eff}$ data obtained during the hydrostatic stage (shown in Figure \ref{Fig:porosity}, see Appendix \ref{AppendixA}) predict that neglecting the porosity change that will occur during the dry stage will lead to an overestimation of the actual sample porosity, or an underestimation of the porosity change, by $\sim 0.6\%$, which is negligible compared to the total porosity change measured during the experiments ($\Delta \phi$ up to $10\%$).

\begin{figure*}
    \centering
    \includegraphics[width=\textwidth]{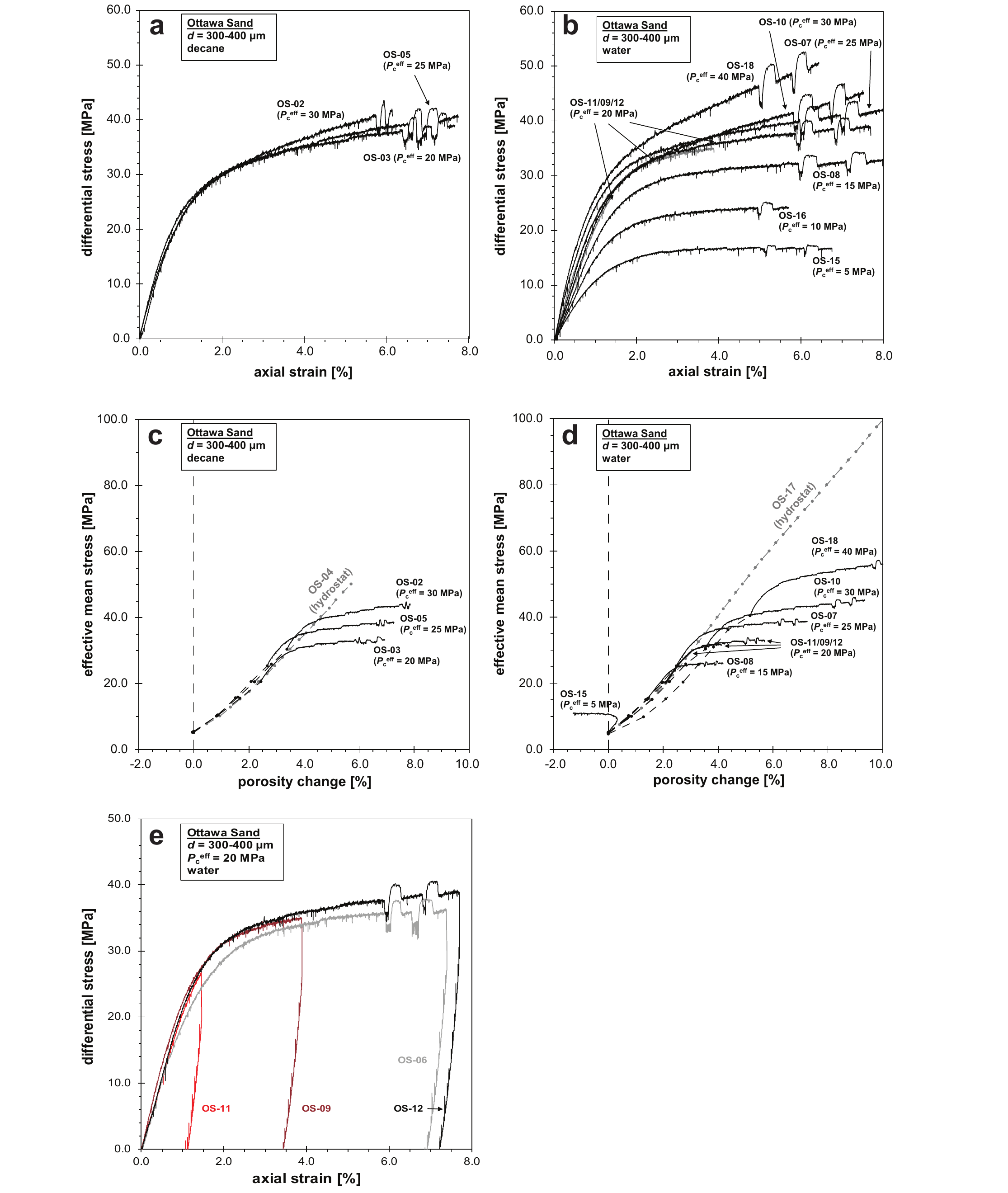}
    \caption{a-b) Differential stress versus axial strain and c-d) effective mean stress versus porosity change curves for triaxial experiments performed on decane-saturated and water-saturated Ottawa sand aggregates, at room temperature and fluid pressures of $5$~MPa. Solid lines indicate triaxial deformation stages. Dotted lines show hydrostatic experiments or experiment stages. Strain rate-stepping stages resulted in up and down steps in differential (a-b) and mean stress (c-d) at the end of each experiment. e) Differential stress versus axial strain curves for selected experiments, illustrating the experimental reproducibility, the impact of porosity on mechanical behaviour (cf. OS--06 and OS--12) and typical unloading behaviour. }
    \label{Fig:Hydrostats}
\end{figure*}

Differential stress-axial strain and mean stress-porosity change curves for hydrostatic and triaxial deformation experiments at effective pressures $P_\mathrm{c}^\mathrm{eff}$ of $5$ to $40$~MPa under decane- and water-saturated conditions (pore-fluid pressure, $P_\mathrm{p}$ = $5$~MPa) are presented in Figure \ref{Fig:Hydrostats}. Overall, all experiments show an initial stage of quasi-linear stress-strain behaviour (Figure \ref{Fig:Hydrostats}a-b). Only at very low effective confining pressure ($P_\mathrm{c}^\mathrm{eff}$ = $5$~MPa; cf. OS-15, Figure \ref{Fig:Hydrostats}d) dilation is observed, while at higher confinement the Ottawa sand aggregates compact. After $\sim$ 1$\%$ strain, the stress-strain curves become non-linear and the samples transition from strain-neutral towards strain-hardening behaviour with increasing effective confining pressure (cf. Figure \ref{Fig:Hydrostats}b). More porosity change is achieved at higher confinement, reaching near-uniaxial compaction (i.e. $\Delta \phi \simeq \epsilon$) at effective confining pressures of $40$~MPa. Note that the slightly lower porosity sample OS--06 showed also slightly lower differential stresses at the same amount of axial deformation (Figure \ref{Fig:Hydrostats}e). Overall, little effect of pore-fluid chemistry (decane vs. water) is observed in terms of aggregate strength and volumetric behaviour. 

The hydrostats (cf. OS-04 and OS-17 in Figure \ref{Fig:Hydrostats}c and d, respectively) show a slightly sigmoidal shape with an initial concave upward trend between mean stress and porosity change, becoming quasi-linear after approximately $2\%$ porosity change and concave downward again after $5\%$ porosity change. At an effective mean stress of $50$~MPa, $5.7~\%$ and $4.9~\%$ porosity change is achieved for the decane- and water-saturated hydrostatic experiment, respectively. The decane hydrostat follows the volumetric behaviour seen in the hydrostatic stages of the decane-saturated triaxial deformation experiments. While the water hydrostat follows the hydrostatic stages of the water-saturated triaxial experiments up to effective confining pressures of $25$~MPa, it underestimates compaction at higher confinement (Figure \ref{Fig:Hydrostats}b). This may be related to sample preparation or to the pressurisation procedure employed during the hydrostatic experiment. It should be noted that within the pressure range tested, the hydrostatic experiments lack any significant inflection point, generally associated with the departure from elastic behaviour and the onset of inelastic compaction ($P^*$ \citep{Zhang1990}).

In our triaxial deformation experiments, the non-linearity of the stress-strain curves during loading makes the determination of a specific onset of inelastic behaviour (commonly denoted $C^{*}$ in sandstones \citep{Wong1997}) very elusive and was therefore not done. Upon unloading, the stress-strain data is near-linear and shows that $80-95 \%$ of the total axial strain is permanent deformation, at all stages of the experiment (Figure \ref{Fig:Hydrostats}e). We use the linear portions of the unloading stress-strain curves to determine the apparent Young's Modulus ($E_\mathrm{u}$ \citep{Zoback1975, BaudSchubnelWong2000} - see Figure \ref{Fig:Hydrostats}e; full data set available from the UK National Geoscience Data Centre). As reported in Table \ref{Tab:DecWat}, Young's Modulus ranges from $3.6$ to $8.1$~GPa in the effective confining pressure range $5$ to $40$~MPa, increasing with $P_\mathrm{c}^\mathrm{eff}$. At constant confining pressure, $E_\mathrm{u}$ remains roughly constant throughout the experiment (cf. OS-11, -09, -06 and -12; Table \ref{Tab:DecWat} and Figure \ref{Fig:Hydrostats}e). Pore-fluid does not affect the static elastic stiffness.

\subsubsection{Acoustical data}
\begin{figure*}
    \centering
    \includegraphics[width=\textwidth]{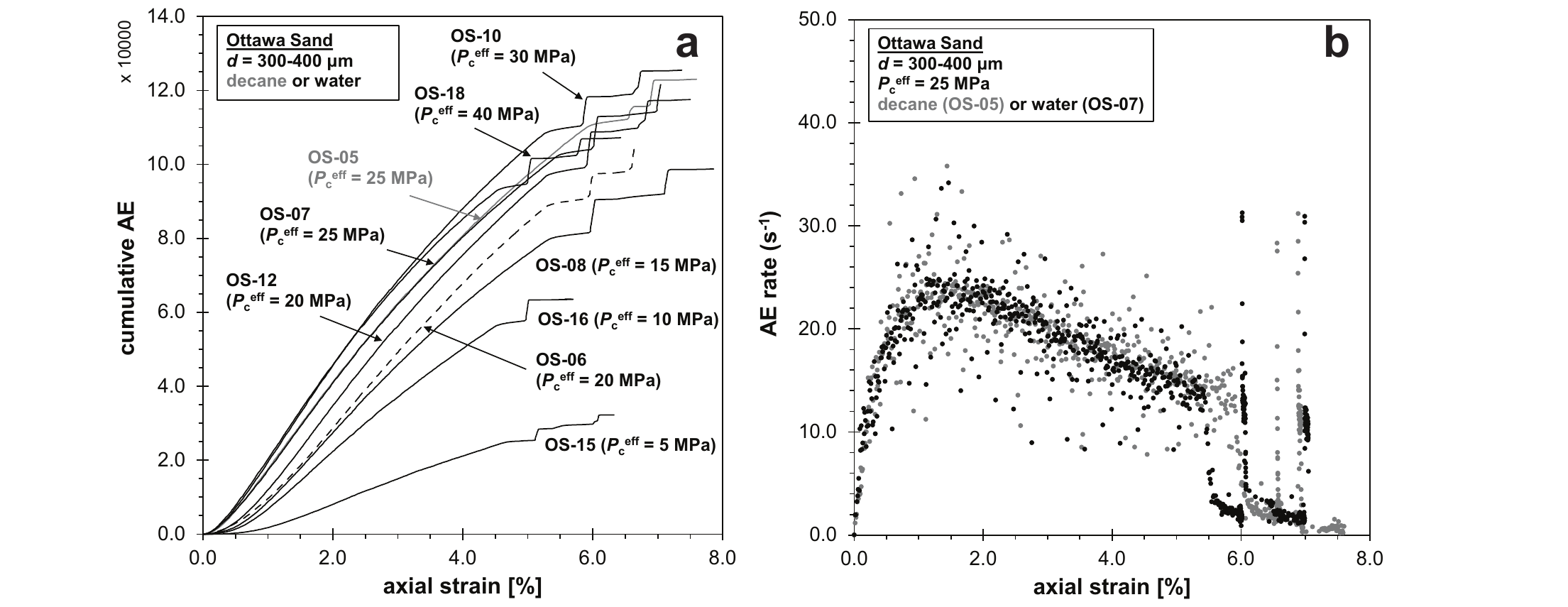}
    \caption{Acoustical data obtained for the decane- and water-saturated experiments ($d$ = $300-400$~$\mu$m). a) Cumulative AE and b) AE rate versus axial strain. Note that the strain rate-stepping stages were initiated after $\sim 5~\%$ axial strain.}
    \label{Fig:AEs}
\end{figure*}

The evolution of the cumulative acoustic emission count and AE rates as a function of axial strain are shown in Figure \ref{Fig:AEs}a and b, respectively. For all experiments, cumulative AEs increase with increasing strain, showing concave upward behaviour up to approximately $1$~\% axial strain followed by concave downward behaviour at larger deformation. Furthermore, more AEs are counted with increasing confining pressure. Note that for experiment OS--06, which had a slightly lower starting porosity, fewer cumulative AEs were counted during compaction.

Acoustic emission rate, i.e. the time derivative of the cumulative AEs, shows a rapid initial increase in rate with axial strain, until it reaches a peak AE rate at an axial strain of $\sim 1\%$, after which the rate slows down again. With increasing confining pressure, the peak AE rate increases in magnitude ($\sim 6.5 ~s^{-1}$ at $5$ ~MPa $P_\mathrm{c}^\mathrm{eff}$ to $\sim 25~s^{-1}$ at $P_\mathrm{c}^\mathrm{eff}$ \textgreater $25$~MPa) and occurs at lower axial strain with increasing confining pressure ($\sim 1.8~\%$ at $5$ ~MPa $P_\mathrm{c}^\mathrm{eff}$ to $\sim 1~\%$ at $P_\mathrm{c}^\mathrm{eff}$ \textgreater $25$~MPa). pore-fluid composition does not affect the cumulative number of AEs or the AE rate (cf. Figure \ref{Fig:AEs}b).

\begin{figure*}
    \centering
    \includegraphics[width=\textwidth]{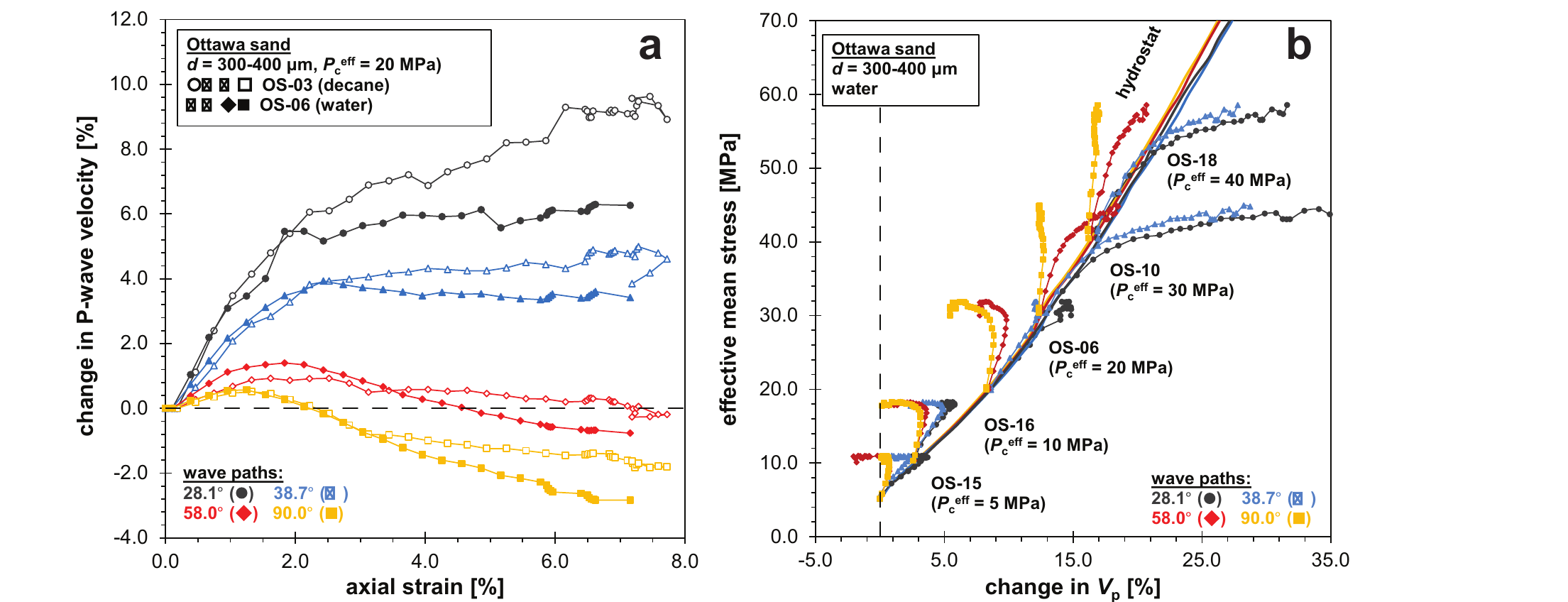}
    \caption{a) Changes in P-wave velocity along the four wave path directions versus axial strain for experiments OS-03 (decane; triangles) and OS-06 (water; circles) at $P_{c}^{eff}$ = $20$~MPa. Strain rate-stepping was initiated at $\sim 5~\%$ axial strain. b) Effective mean stress versus change in P-wave velocity at different effective confining pressures for water-saturated Ottawa sand aggregates. The triaxial deformation experiments are superimposed onto the hydrostat. Note that strain rate-stepping stages were initiated near the end of each experiment.}
   \label{Fig:Wave-vel}
 \end{figure*}

Absolute P-wave velocities during the decane- and water-saturated experiments, measured along the four wave paths, range from $1800$ to $2250$~m/s prior to deformation, increasing with increasing confinement, i.e. with decreasing porosity at the start of the triaxial deformation stage. The velocities measured through the lower half of the samples were consistently higher, most likely due to a slightly denser packing of the aggregate resulting from the preparation method. Changes in P-wave velocity with respect to the pre-deformation value are shown in Figure \ref{Fig:Wave-vel}a as a function of axial strain for decane- and water saturated sand ($P_\mathrm{c}^\mathrm{eff}$ = $20$~MPa). For all experiments, P-wave velocity increases rapidly for the initial $1-2\%$ strain. At low confining pressure ($P_\mathrm{c}^\mathrm{eff}$ \textless $20$~MPa), P-wave velocity decreases along all four wave paths, and eventually becomes less than the starting value for the (sub-)horizontal wave paths ($58.0~^{\circ}$ and $90.0~^{\circ}$). By contrast, at higher confining pressure, after the initial rapid increase, P-wave velocity stays constant or increases at a slower rate along the steeper wave paths ($28.1~^{\circ}$ and $38.7~^{\circ}$, and  $58.0~^{\circ}$ at the highest $P_\mathrm{c}^\mathrm{eff}$). Upon unloading P-wave velocity remains nearly constant for all experiments, suggesting that most of the measured deformation is permanent in line with the mechanical data (cf. Section \ref{Mech_data}). The evolution of $V_\mathrm{p}$ is not significantly affected by the pore-fluid type (Figure \ref{Fig:Wave-vel}a).

For water-saturated aggregates, the evolution of P-wave velocity with effective mean stress, superimposed onto the hydrostat, is shown in Figure \ref{Fig:Wave-vel}b. For the selected experiments, changes in P-wave velocity are shown along all four wave paths, to illustrate the development of azimuthal anisotropy of damage as a function of confining pressure. For all triaxial deformation experiments, the change in P-wave velocity initially increases linearly with effective mean stress and shows similar or slightly lower values compared to those measured under hydrostatic conditions. The initial rate of P-wave velocity increase is the lowest for the horizontal wave path, increasing as wave path angle rotates to sub-vertical, though no dependence on confinement is seen. As effective mean stress increases P-wave velocity decreases with increasing mean stress along all wave paths for effective confining pressures up to $30$~MPa. By contrast, at high confinement ($P_\mathrm{c}^\mathrm{eff} \geq 30$~MPa) P-wave velocities increase rapidly with increasing stress, and thus strain, except along the horizontal wave path. The increase in wave velocity is most rapid in the sub-vertical direction, i.e. sub-parallel to the compression direction. All experiments display an increase in wave-anisotropy (i.e. $V_p^{28.1~^{\circ}}/V_p^{90.0~^{\circ}}$) due to deformation, which becomes stronger with higher confinement (see Figure \ref{Fig:Wave-vel}b). At low $P_\mathrm{c}^\mathrm{eff}$, anisotropy is slightly above one, at $1.05$ to $1.1$. As confining pressure increases to $30-40$~MPa, anisotropy increases to $1.2-1.3$.

\subsubsection{Microstructural analysis and located AE events}
\begin{figure*}[t]
    \centering
    \includegraphics[width=0.95\textwidth]{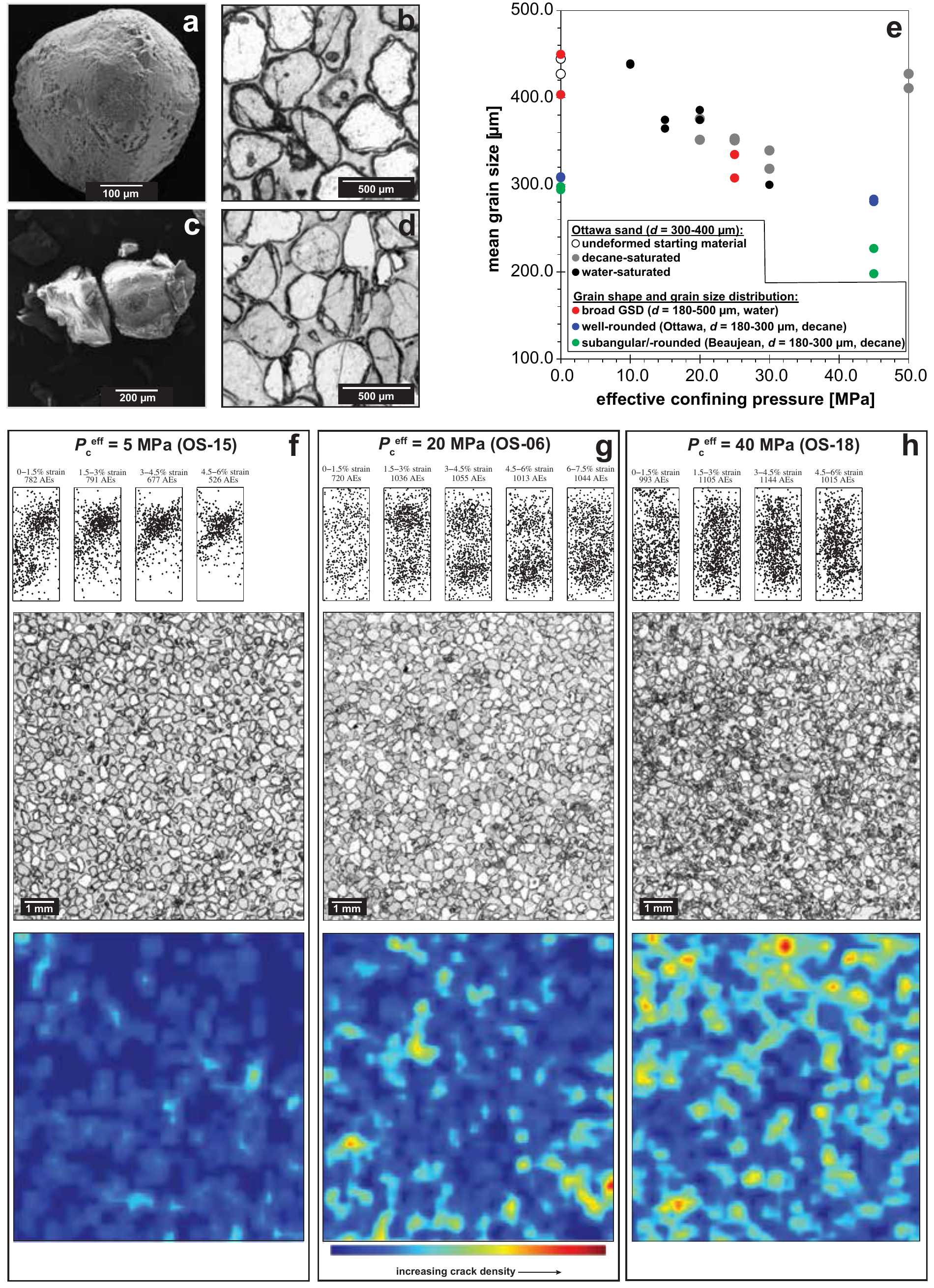}
    \caption{Scanning Electron and optical micrographs of Ottawa sand grains, which are a) undeformed, b) compacted at low confining pressure (OS-15; $P_{c}^{eff}$ = $5$~MPa) and c-d) compacted at high confining pressure (OS-12 (c) and OS-06 (d); $P_{c}^{eff}$ = $20$~MPa). e) Mean grain-size ($d_\mathrm{50}$) versus confining pressure, showing a reduction in mean grain-size with increasing confinement, indicative of the chipping and fracturing seen in (b-d). Located AE events for the whole sample (40 by 100 mm), optical micrographs of sections taken at the centre of the sample (10 by 10 mm) and corresponding crack density contour maps obtained from water-saturated Ottawa sand aggregates, deformed at e) $5$~MPa (OS-15), f) $20$~MPa (OS-06) and g) $40$~MPa (OS-18) effective confining pressure. Note that the micrographs are not taken from the same slice through the sample as the AE locations.} 
    \label{Fig:TSA+GSA}
  \end{figure*}

Undeformed Ottawa sand ($d$ = $300-400$~$\mu$m) consists of well-rounded grains with relative smooth surfaces (Figure \ref{Fig:TSA+GSA}a). After deformation, the grains have a more ragged appearance, most likely due to grain edge chipping such as seen in Figure \ref{Fig:TSA+GSA}b. Chipping is evident at low confinement (Figure \ref{Fig:TSA+GSA}b), where little grain crushing occurs. At higher confinement, fracturing and crushing becomes more dominant, shattering grains into angular fragments and small angular flakes (Figure \ref{Fig:TSA+GSA}c and d). These microstructural observations are confirmed by grain-size analyses (Figure \ref{Fig:TSA+GSA}e), which show the increased formation of finer particulates with increasing confinement. Very little grain-size reduction is observed in hydrostatic experiment OS-04 (decane), though compaction is following the same trend as the hydrostatic phases of the triaxial deformation experiments. No grain-size analysis could be performed on the water-saturated hydrostatic experiment (OS-17).

AE events indicate that localisation at low confining pressure ($P_\mathrm{c}^\mathrm{eff}$ $\leq$ $5$~MPa) occurs along a shear plane. Though not evident from the microstructure, the volumetric data suggests that this plane is likely a dilation band. In line with this, pervasive grain failure is absent in the microstructure (Figure \ref{Fig:TSA+GSA}e), while grain edge chipping is prevalent, facilitating the required grain rotation and rearrangement needed for strain localisation. At higher confining pressures ($P_\mathrm{c}^\mathrm{eff}$ $\geq$ $10$~MPa) deformation appears to be more diffuse, with no clear indication of localisation. Furthermore, the degree of grain fracturing and crushing increases with axial strain from $5.3$ cracks/mm$^2$ at $1.5\%$ strain (OS-09) to $5.6$ cracks/mm$^2$ at $4.5\%$ strain (OS-11) and $11.4$ cracks/mm$^2$ at $7.5\%$ strain (OS-06; $P_\mathrm{c}^\mathrm{eff}$ = $20$~MPa), as well as with confining pressure (e.g. from $5.0$ cracks/mm$^2$ at $5$~MPa to $16.9$ cracks/mm$^2$ at $40$~MPa $P_\mathrm{c}^\mathrm{eff}$). The microstructures do not show clear indications for strain localisation (cf. Figure \ref{Fig:TSA+GSA}f-h).

\subsection{Effect of strain rate}
After $\sim 5\%$ axial strain, strain rate-stepping stages were performed by stepping the strain rate by one order of magnitude down and up. During strain rate down-steps the differential stress, and hence also the mean stress, measured across the sample decreased. Similarly for strain up-steps the differential and mean stress increased (cf. Figure \ref{Fig:Hydrostats}). The magnitude of these stress changes appears to increase with increasing confining pressure.

The strain rate steps are also clearly reflected in the cumulative acoustic emission counts and AE rates. Decreasing the strain rate significantly reduces the AEs produced by the sample, down to a rate of \textless $3$~$s^{-1}$. Conversely, increasing the strain rate resulted in a rapid increase in the number of AEs, to a rate of $\sim 30$ $~s^{-1}$ (Figure \ref{Fig:AEs}). However, P-wave velocities are not significantly impacted by the strain rate-stepping sequence, which is in accordance with the observation that strain rate-stepping does not affect sample volume changes.

\subsection{Experiments varying grain angularity and grain-size distribution}
In addition to the experiments on well-rounded Ottawa sand with a narrow grain-size, we also performed experiments on sub-rounded/sub-angular Beaujean sand ($d$ = $240 \pm 60$~$\mu$m) and Ottawa sand with a broader grain-size distribution ($d$ = $340 \pm 160$~$\mu$m). These experiments aimed at studying the effect of grain angularity (OS-19 and BS-01, $P_\mathrm{c}^\mathrm{eff}$ = 45~MPa, decane) and grain-size distribution (OS-07 and OS-14, $P_\mathrm{c}^\mathrm{eff}$ = 25~MPa, water) on compaction behaviour (Table \ref{Tab:GSD-shape}). 

\begin{figure}
    \centering
    \includegraphics[width=\linewidth]{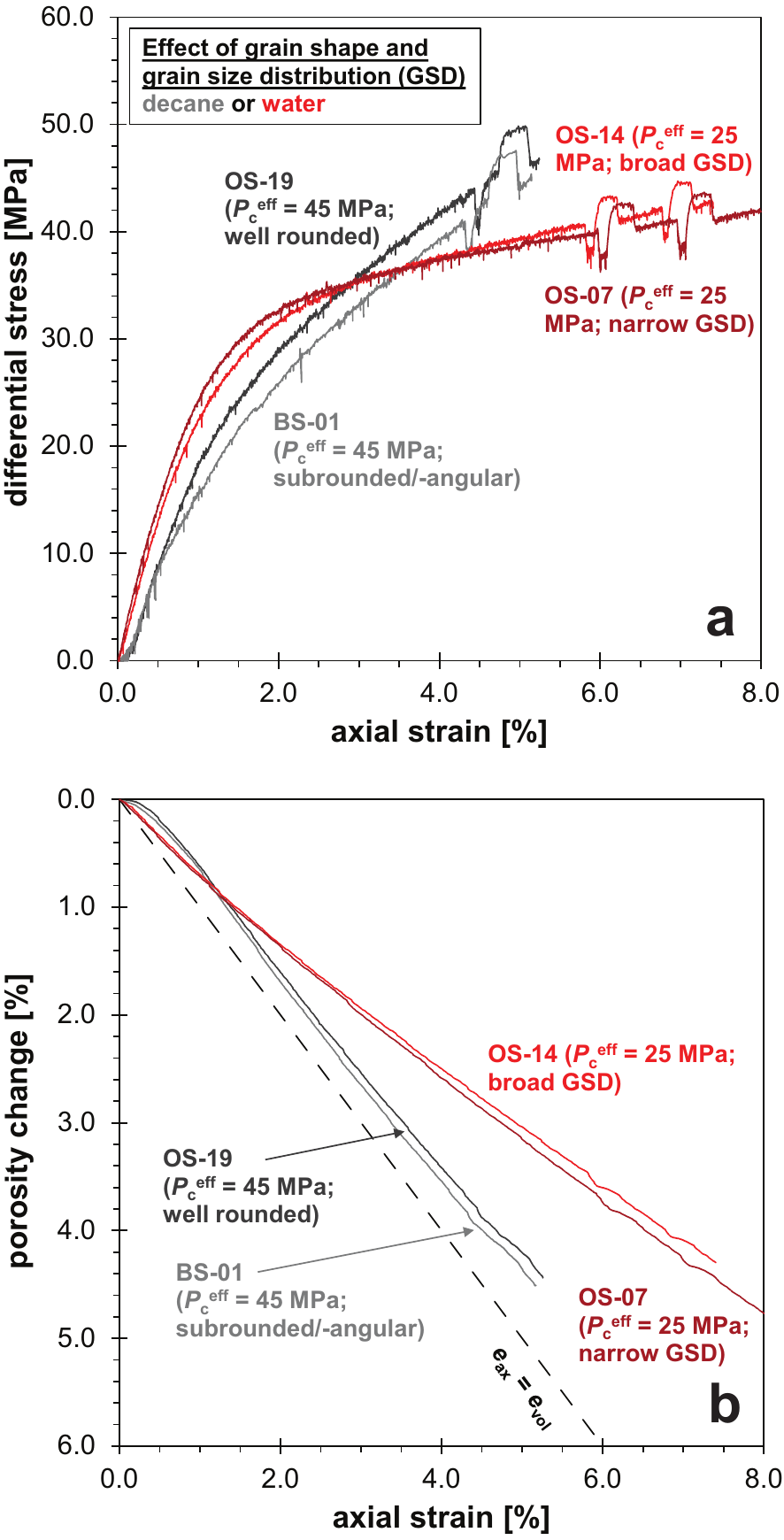}
    \caption{a) Differential stress versus axial strain and b) porosity change versus axial strain curves for triaxial experiments performed to test the effect of grain angularity (grey curves) and grain-size distribution (GSD - red curves) on mechanical behaviour. Note that experiments OS-19 and BS-01 were performed on decane-saturated aggregates, while OS-07 and OS-14 were performed on water-saturated aggregates, at room temperature and fluid pressures of $5$~MPa. Strain rate-stepping stages resulted in up and down steps in differential stress at the end of each experiment.}
    \label{Fig:GSD-shape}
\end{figure}

\begin{figure}
    \centering
    \includegraphics[width=\linewidth]{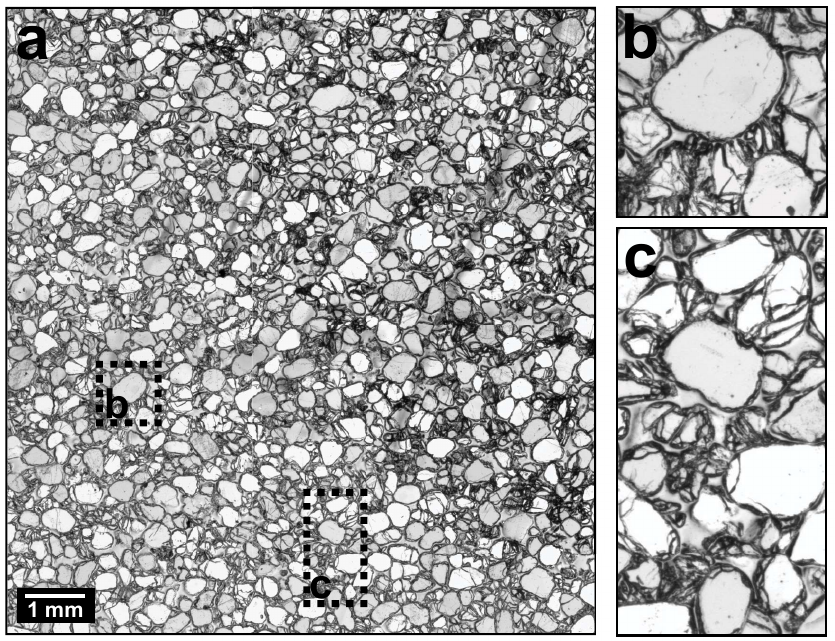}
    \caption{a) Optical micrograph of Ottawa sand aggregates with a broad grain-size distribution ($d$ = $180-500$ $\mu$m), deformed at $25$~MPa (OS-14), with b) and c) insets showing large surviving grains surrounded by crushed particles, suggesting shielding of the coarser grains by the finer grains present within the grain-size fraction.}
    \label{Fig:TSA_GSD}
\end{figure}

Differential stress-axial strain and porosity change-axial strain curves are presented in Figure \ref{Fig:GSD-shape}. All four aggregates show behaviour similar to the Ottawa sand described in Section \ref{main-OS-expts} with quasi-linear loading and near-uniaxial compaction behaviour. All four experiments show significant strain-hardening, which is more pronounced at high confining pressure. In terms of the effect of grain angularity, the mechanical data suggest that there is no significant difference in behaviour between the well-rounded Ottawa sand and the more subrounded/-angular Beaujean sand. It is evident from Malvern grain-size analysis that the Beaujean sand underwent significant grain crushing, with $d_\mathrm{50}$ values reducing by $\sim 28~\%$ from $296$~$\mu$m to $212$~$\mu$m (compared to only $\sim 8\%$ reduction for Ottawa sand, from $308$~$\mu$m to $282$~$\mu$m). Similarly, varying grain-size distribution, i.e. narrow vs. broad, does not affect the mechanical behaviour of the Ottawa sand aggregates. Microstructural observations show that the coarser grains within the aggregate remain largely intact during deformation, while the finer grains have been crushed (Figure \ref{Fig:TSA_GSD}). Compared to the narrow GSD Ottawa sand deformed at $25$~MPa (OS-05), the average grain-size has been significantly reduced, as a result of the formation of fines ($d_\mathrm{50}$ is $321$~$\mu$m and $352$~$\mu$m, respectively; see Figure \ref{Fig:TSA+GSA}e). No indications for strain localisation are observed.

\section{Discussion}
\subsection{Strain localisation}
Preservation of deformation structures in sands is challenging due to the unconsolidated nature of the material. However, shear bands in sands have been recognised in nature \citep[e.g. see][]{Cashman2000, Rawling2003} and in the laboratory \citep{Alikarami2015}. Structures seen in the Aztec sandstone are inferred to suggest that compaction bands may also form in unconsolidated sands \citep{Sternlof2005}, though to date no such laboratory observations have been made. By contrast, discrete compaction bands have been observed in sandstones in both nature \citep{Mollema1996, Sternlof2006} and in the laboratory \citep[see][for an overview]{Wong2012}. While there may be differences in their characteristics, both natural and lab-induced compaction bands indeed show locally intense microcracking, grain comminution and porosity reduction. However, not all sandstones are prone to compaction band formation. Porosity and rock homogeneity (mineralogy, grain-size distribution) play a significant role in controlling this type of strain localisation \citep{Cheung2012, Mollema1996}, though their interplay is not yet well understood. Our results suggest that there is no evidence for compaction localisation in unconsolidated sand deformed in the compactant regime.

\subsection{Effect of water, grain angularity and grain-size distribution on behaviour}
In the presence of chemically active fluids like water, sandstones generally tend be significantly weaker \citep{Baud2000, Duda2012, Baud2015, Tembe2008}. Such physico-chemical weakening of sandstones is attributed to the presence of altered feldspars and clay minerals \citep{Baud2000}, which reduce the proportion of strong, quartz-cemented grain-to-grain contacts. Furthermore, altered feldspars may promote crack nucleation and growth, while clays may promote slip through a reduction in frictional resistance when wet \citep{Reviron2009, Baud2000}. By contrast, very pure sandstones, like Fontainebleau and Bentheim sandstone, show no significant water-weakening \citep{Reviron2009, Tembe2008}. Similarly, our experiments on decane- and water-saturated quartz sand ($d$ = $300-400$~$\mu$m) show no clear effect of a chemically active fluid on stress-strain and strain-$V_p$ behaviour, and grain-size reduction within the range of strain rates investigated (cf. Figures \ref{Fig:Hydrostats}, \ref{Fig:Wave-vel} and \ref{Fig:TSA+GSA}). These observations are in accordance with the behaviour seen during time-independent uniaxial compaction of pure quartz sand, which is not significantly affected by chemical environment \citep{Brzeso2014}. This underpins the suggestion that the presence of other mineralogical phases is needed to promote water-weakening. Recent work on impure sandstone inferred that intergranular slip at clay-coated grain boundaries may play an important role in controlling deformation, especially at stress conditions below the yield point \citep{Pijnenburg2018, Pijnenburg2019}.

Grain angularity is known to lead to sharper contact points and hence higher contact stresses, leading to more grain-breakage \citep{Chuhan2003, Zhang1990}, i.e. the potential to create an instability. At the same time, in sandstones with a broad grain-size distribution, large grains tend to inhibit the propagation of instabilities, thereby hindering the formation of compaction bands \citep{Cheung2012}. We performed two experiments aimed at investigating the effect of grain angularity (Ottawa sand OS-19 vs. Beaujean sand BS-01, $P_\mathrm{c}^\mathrm{eff}$ = $45$~MPa, decane). Previous triaxial experiments on Ottawa and Hostun sand under lower confining conditions ($P_\mathrm{c}$ = $0.1-7$~MPa; see \citet{Alikarami2015}) showed comparable stress-strain behaviour for the two sands, though the angular Hostun sand displayed more pronounced dilatation and grain-breakage. While our experiments are performed at higher confining pressure conditions, we also observe very similar stress-strain behaviour between the Ottawa and Beaujean sand. Furthermore, the more angular Beaujean sand shows significant grain failure and a smaller average grain-size compared to the rounded Ottawa sand. We speculate that the more angular Beaujean sand has a higher tendency to strong interlocking, which would require the 'chipping off' of larger fragments to unlock the aggregate and promote grain rearrangement, compared to the smoother, well-rounded Ottawa sand grains.

It should also be noted that overall finer-grained Ottawa sand ($d$ = $180-300$~$\mu$m) showed less grain-size reduction than coarser-grained Ottawa sand ($d$ = $300-400$~$\mu$m), even when deformed at higher effective confining pressure. This is inferred to be related to the tendency for larger grains to have a lower yield stress, making them more prone to grain failure \citep{Chuhan2003, Zhang1990}. Regarding grain-size distribution, stress-strain behaviour is very comparable between the two end-members (OS-07 vs. OS-14, $P_\mathrm{c}^\mathrm{eff}$ = $25$~MPa, water). Looking at the microstructures, the broad GSD experiment displays shielding of the larger grains by the smaller grains, leading to less breakage of the coarser grains, supporting the observations made for Boise sandstone \citep{Cheung2012}. Overall, our experiments on Ottawa sand show that in terms of chemical environment, grain angularity and grain-size distribution, sands tend to show qualitatively similar behaviour compared to sandstones.

\subsection{Hardening envelopes for sand in PQ-space} \label{hardening}
\begin{figure*}
    \centering
    \includegraphics[scale=0.75]{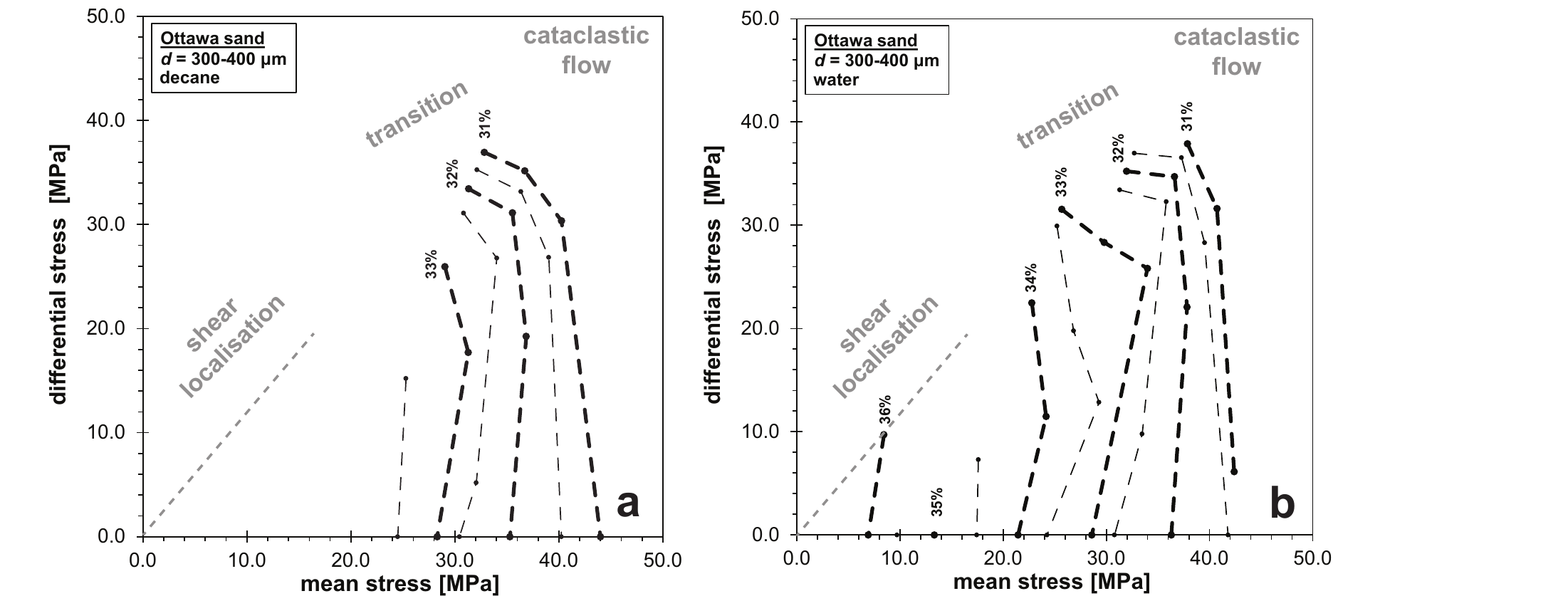}
    \caption{Hardening envelopes for specific aggregate porosity-values plotted in differential stress versus mean stress space for a) decane-saturated and b) water-saturated Ottawa sand ($d$ = $300-400\mu$m) aggregates, at room temperature and fluid pressures of $5$~MPa. Thick and thin dashed black lines separate porosity changes of $0.5\%$.}
    \label{Fig:PQcontours}
\end{figure*}

For sandstones, the yield point is typically defined as the stress conditions at which the material is no longer behaving poro-elastically \citep{Wong2012}. This point is generally accompanied with a marked increase in acoustic emission activity \citep[e.g.][]{Tembe2007, Fortin2006}. However, in the case of unconsolidated sand, large permanent deformation is seen during loading from the onset and elastic behaviour is only observed upon unloading (Figure \ref{Fig:Hydrostats}e). As such, no clear yield point can be pinpointed on the basis of the mechanical data, the acoustic emission activity or the P-wave velocities (cf. Figures \ref{Fig:Hydrostats}b, \ref{Fig:AEs}a and \ref{Fig:Wave-vel}b). Therefore, to generate a meaningful description of the deformation behaviour of this material, we propose it is more appropriate to represent the stress-strain data in terms of expanding hardening envelopes, each representing the stress supported at a specific porosity value. Since the sand aggregates continuously compact, this means that the hardening envelopes gradually grow as aggregate porosity decreases \citep{Karner2005}. A single loading path through these hardening envelopes then gives the stress-strain behaviour of the material. As the mechanical behaviour of Ottawa sand is reproducible and its material properties are near-constant throughout each experiment, delineating the mechanical behaviour in terms of hardening envelopes is warranted (cf. compare OS-09, -11 and -12, see Table \ref{Tab:DecWat} and Figure \ref{Fig:Hydrostats}e). For both the decane- and water-saturated aggregates, these hardening envelopes are drawn in differential stress, $Q = (\sigma_1 - \sigma_3)$, versus mean stress, $P = (\sigma_1 + 2\sigma_3)/3 - P_\mathrm{p}$, space (Figure \ref{Fig:PQcontours}).

As is clear from the hardening envelopes, Ottawa sand shows significant porosity reduction during the hydrostatic stage of the experiment (up to $\sim 5\%$ at $40$~MPa $P_\mathrm{c}^\mathrm{eff}$, cf. Figure \ref{Fig:Hydrostats}), which has also been observed for St. Peter sand \citep{Karner2003}. However, the most striking feature of the contours is that they are near-vertical up to $\sim30$~MPa differential stress, suggesting that porosity change is mainly controlled by mean stress, and not significantly by differential stress. Similar behaviour has been observed before in St. Peter sand aggregates \citep{Karner2005}, porous bassanite \citep{Bedford2018} and even in some highly porous sandstones \citep{Tembe2007, Pijnenburg2019}. 

In essence, the hardening envelope for sand appears to consist of three linear parts: a shear failure line at low $P$, a narrow mean stress-range during which deformation appears to be pressure-insensitive, followed by a rapid transition to a near-vertical end cap, which increases in size as porosity decreases (see Figure \ref{Fig:PQcontours}). The bifurcation analysis of \citet{Issen2000} predicts that localised compaction bands form when the sum of the dilatancy factor $\beta$ (the ratio of inelastic volumetric over shear strain increments) and friction parameter $\mu$ is $(\beta+\mu)<-\sqrt{3}$. Analysis of our data to tease out these constitutive parameters (see Table \ref{Tab:ConstParam}) suggests that at least in the highest confining pressure experiments conditions were favourable to promote compaction band formation. However, this prediction is at odds with our observations. Such a discrepancy between the prediction from bifurcation theory and experimental observations is not uncommon for sandstones, and is likely due to the inadequacy of the constitutive description \citep{Baud2006}. Furthermore, it should be noted that we do not have access to the yield envelope \textit{sensu stricto}, but to hardening envelopes, which might develop a different slope from the conventional yield envelope. This also inhibits a meaningful comparison between sand and sandstones. Given this discrepancy in behaviour between sand and sandstone, the question now is: what are the underlying microphysical processes controlling this behaviour?

\begin{table*}
    \caption{Constitutive parameters obtained for conventional triaxial compression experiments performed on water-saturated Ottawa sand aggregates ($d = 300-400 \mu$m) at $20 ^\circ$C and $P_\mathrm{p}$ = $5$~MPa.}
    \label{Tab:ConstParam}
    \centering
    \begin{tabular}{l c c c c c c c c c c}
    \hline
 Sample                         & $P_\mathrm{c}^\mathrm{eff}$    & $\beta_\mathrm{33\%}$      & $\mu_\mathrm{33\%}$        & $h_\mathrm{tan, 33\%}$        & $\beta_\mathrm{32\%}$      & $\mu_\mathrm{32\%}$        & $h_\mathrm{tan, 32\%}$        & $\beta_\mathrm{31\%}$      & $\mu_\mathrm{31\%}$        & $h_\mathrm{tan, 31\%}$ \\
                                & [MPa]         & [-]          & [-] & [GPa]          & [-]          & [-]          & [GPa]          & [-]          & [-]          & [GPa]  \\
\hline
\hline
  OS--08                        & 15             & 0.13            & 1.83            & 0.01            & 0.19            & 0.67            & 0.10            &             &             & \\
  OS--09\tablenotemark{a}       & 20            & 0.23            & 0.86            & 0.80            & 0.20            & 0.31            & 0.08            &             &             & \\
  OS--12                        & 20            & 0.27            & 0.92            & 0.87            & 0.23            & 0.16            & 0.09            &             &             & \\
  OS--07                        & 25            & 0.39            & -0.30            & 1.19            & 0.30            & -0.82            & 0.37            & 0.29            & -1.02            & 0.09\\
  OS--10                        & 30            & 0.39            & -0.30            & 1.19            & 0.38            & -1.73            & 1.53            & 0.33            & -1.93            & 0.47\\
  OS--18                        & 40            &             &             &             &             &            &             & 0.35            & -12.11            & 3.88\\
  $ $ \\
\hline
\end{tabular}
\tablenotetext{}{Symbols: $P_\mathrm{c}^\mathrm{eff}$ represents effective confining pressure, $\beta$, $\mu$ and $h_\mathrm{tan}$ are the constitutive parameters as needed for bifurcation analysis \citep{Rudnicki1975, Issen2000}, obtained at sample porosities of 33\%, 32\% and 31\%, respectively. Here $\beta$ is the dilatancy factor, $\mu$ is the internal friction coefficient and $h_\mathrm{tan}$ is the tangent hardening modulus, obtained from the slope of a differential stress-axial strain plot (see Figure  \ref{Fig:Hydrostats}b).}
\tablenotetext{a}{Sample deformed up to 4$\%$ $e$.}
\end{table*}

\subsection{Mechanisms driving compaction and potential for strain localisation in sands}
In our lowest confinement experiment ($P_\mathrm{c}^\mathrm{eff}$ = $5$~MPa), we observed a dilating shear band, based on AE locations (cf. Figure \ref{Fig:TSA+GSA}f), similar to what is observed in Ottawa sand experiments under similar conditions using X-ray tomography imaging \citep{Alikarami2015}. However, no strain localisation was observed at higher confinement. If strain localisation had occurred, this would have been evident in the form of stress drops in the stress-strain behaviour (Figure \ref{Fig:Hydrostats}; \citet{Baud2004}), or visible in the AE locations (Figure \ref{Fig:TSA+GSA}g-h; \citet{Fortin2006, Olsson2000}), or crack density variations within the samples (crack density contour maps in Figure \ref{Fig:TSA+GSA}g-h; \citet{Baud2004, Fortin2006}). Instead, under these conditions cataclastic flow appeared to control deformation, as also suggested by the near-vertical shape of our hardening envelopes (see Section \ref{hardening}).

As summarised by \citet{Rudnicki2007}, compaction bands grow due to high stress concentrations at their tips. In other words, the failure of a grain should lead to the grain no longer supporting any stress and the subsequent redistribution of the stress to the surrounding grains, i.e., it needs to lead to strain-weakening/-softening. Throughout all experiments, Ottawa sand shows near-strain-neutral to strain-hardening behaviour (Figure \ref{Fig:Hydrostats}a and b, Table \ref{Tab:DecWat}). The progressive hardening of the material is also evidenced by the AE locations, which appear to propagate from the sample edge towards the centre, and the continuously increasing P-wave velocities, suggesting a stronger, denser aggregate is being formed (cf. Figure \ref{Fig:Wave-vel}). Therefore, the condition required for compaction band propagation, the creation of an instability, does not appear to occur.

Instead, we speculate that our data suggest that even at high confining pressure, grain-breakage, and concomitant grain rotation and rearrangement controls compaction. It is difficult to differentiate between AE events emerging from grain-breakage and grain rotation \citep{Wong1992}, meaning that the acoustic emission locations could be derived from either (see Figure \ref{Fig:TSA+GSA}f-h), as also suggested by \citet{Menendez1996}. However, significantly fewer AEs are observed at low confinement (Figure \ref{Fig:AEs}a), compared to high $P_\mathrm{c}^\mathrm{eff}$ experiments. Microstructural evidence suggests that grain chipping (Figure \ref{Fig:TSA+GSA}b) is prevalent at these low confinement conditions, while progressively more grain-breakage occurs as confinement increases (see crack density contour maps in Figure \ref{Fig:TSA+GSA}f-h). 

\begin{figure*}
    \centering
    \includegraphics[scale=.75]{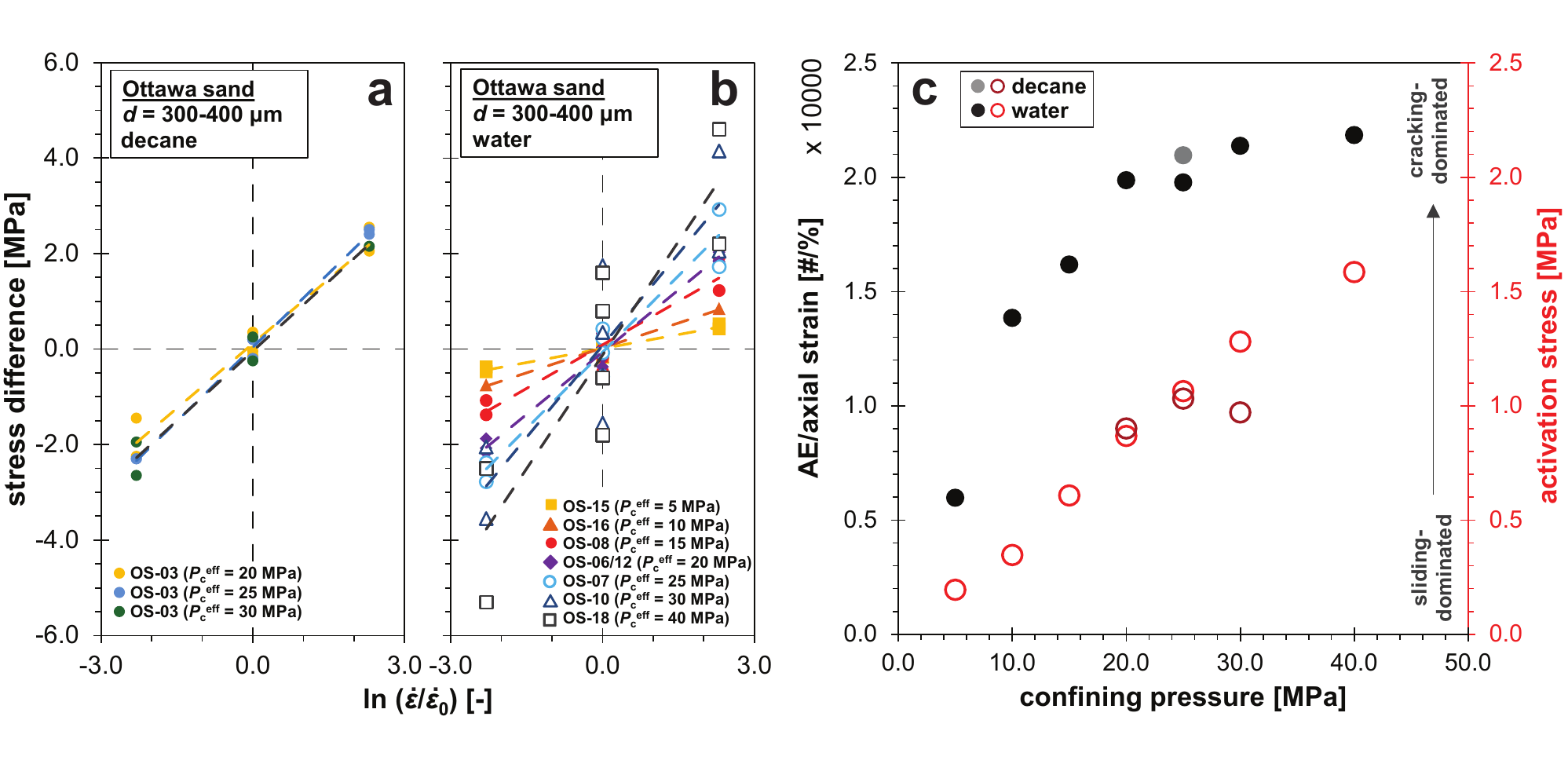}
    \caption{Differential stress difference measured upon a change in axial strain rate versus normalised strain rate for a) decane-saturated and b) water-saturated aggregates. c) Cumulative number of AEs per unit axial strain and activation energy versus effective confining pressure for decane- and water-saturated aggregates.}
    \label{Fig:SR-dependence}
\end{figure*}

This change in microstructural behaviour with confining pressure is also suggested from the strain rate stepping data. As shown in Figures \ref{Fig:SR-dependence}a and b, upon a change in axial strain rate, there is a change in differential stress supported by the sample. Overall, the differential stress change increases in magnitude when confining pressure increases, i.e. when aggregate porosity goes down. Following \citet{Brantut2014}, given that the strain rate is related to the differential stress as $\sigma^*$ ln($\dot{\epsilon}$/$\dot{\epsilon_0}$) $\approx$ $\Delta Q$, where $\sigma^*$ is the activation stress, shows that at low $P_\mathrm{c}^\mathrm{eff}$ (high porosity), the activation stress is very low ($<$ $0.5$ ~MPa), while at higher confinement (lower porosity) the value is high ($>$ $1$ ~MPa). High activation stresses are generally associated with cataclastic behaviour in sandstones, e.g. $\sigma^*$ = $4-5$~MPa for compaction band formation in Bleurswiller sandstone \citep{Heap2015}. Similarly, looking at the cumulative AE data, we can see that fewer AEs are produced per unit axial strain at low confining pressure. The rate of AEs with unit axial strain (AE efficiency, see \citet{Wong1992}) increases with confinement (Figure \ref{Fig:SR-dependence}c). It has been suggested that this is generally associated with a transition from compaction by predominantly grain rotation/rearrangement, associated with minor grain fracturing (chipping), to predominantly grain failure plus grain rotation \citep{Wong1992}, in line with the observed strain rate dependency.

The above hypothesis for the grain-scale behaviour is supported by the wave velocity data, which do not show significant changes in horizontal $V_p$, indicative of the opening of axial cracks. Furthermore, the decrease in P-wave velocities as failure approaches, such as observed for sandstones reflecting enhanced microcracking and pore collapse \citep{Fortin2006}, is also not observed. Though our samples experience significant grain failure, as evidenced by the grain-size and thin section analyses, the P-wave velocity appears not to be controlled by the formation of cracks. Instead, it is inferred that the data reflect that the acoustic behaviour of Ottawa sand is associated with the contact area between grains \citep{Digby1981, Walton1987}.

Overall, we speculate that for our sand aggregates grain failure is easily accommodated by the rearrangement of surrounding grains, thereby preventing stress concentrations to arise within the aggregate. As the aggregate is subjected to higher confining pressures, it achieves a lower porosity and a more locked microstructure. Furthermore, higher confinement promotes more lateral constraint. So, while grain chipping is sufficient to promote rearrangement at low $P_\mathrm{c}^\mathrm{eff}$, grain crushing becomes more important at higher $P_\mathrm{c}^\mathrm{eff}$, as the grains experience less freedom to move around, as reflected in our microstructural, AE and strain rate data. However, even at high confinement, grain rearrangement is still possible meaning that no stress concentration, and hence no strain localisation, will occur within the aggregate. The potential for compaction band formation therefore seems to be hinging on the degree of cementation at the grain contact, though it is unclear what the minimum degree of cementation should be and more research is needed on this \citep[e.g.][]{Bernabe1992, Yin1994}.

%
%

\section{Conclusions}
To probe the key differences between the mechanical behaviour of unconsolidated sand and cemented porous sandstones, we performed systematic triaxial compression experiments on well-rounded Ottawa quartz sand at room temperature, strain rate of $10^{-6}-10^{-4}~s^{-1}$ and effective confining pressures of $5-40$~MPa. All samples had a well controlled starting porosity of 36.0-36.3$\%$, a narrow grain-size distribution ($d$ = $300-400$ $\mu$m) and were either decane (inert fluid) or water (chemically active fluid) saturated. Additionally, a limited number of experiments were performed to assess the effect of grain angularity (Beaujean sand) and a broader grain-size distribution ($d$ = $180-500 \mu$m) on behaviour. The observed mechanical, acoustical and microstructural behaviour was compared to that seen in highly porous sandstones, in order to elucidate whether strain localisation, in the form of shear and compaction bands, can occur in sands as well. We observed the following:
\begin{enumerate}
   \item Pure quartz Ottawa sand initially shows quasi-linear stress-strain behaviour going to strain-neutral and strain-hardening behaviour with increasing effective confining pressure. Overall, only at very low $P_\mathrm{c}^\mathrm{eff}$ ($5$~MPa) dilatant behaviour is seen, while compacting increases with effective confining pressure. This behaviour is also reflected in the P-wave velocities, with initial $V_p$-values of $1800-2300$~m/s that increase by up to $20 \%$ as compacting progresses.
   \item The presence of an inert or chemically active fluid did not significantly affect the mechanical and acoustical behaviour of Ottawa sand. These observations are in line with what is observed for pure quartz sandstones.
   \item Grain angularity and grain-size distribution also did not affect the stress-strain behaviour of the sand. More angular Beaujean quartz sand shows more grain crushing, which is expected due to the presence of sharper grain-to-grain contacts, leading to higher contact stresses. For aggregates with a broader grain-size distribution, large grains tend to be shielded by smaller grains, which end up being crushed. Qualitatively this behaviour is also seen in sandstones, such as the Boise sandstone.
   \item As our sand aggregates do not show truly elastic and yield behaviour, the mechanical behaviour is better described using hardening envelopes. In PQ-space, these hardening envelopes appear to consist of a linear shear failure line and a rapid transition to a near-vertical end cap. At low confining pressure, we see the formation of a dilation shear band, while at high confinement cataclastic flow dominates compaction behaviour, in line with the behaviour expected from the hardening envelopes. No compaction bands are observed.
   \item The lack of cementation is inferred to lead to relatively easy grain rotation and rearrangement upon grain chipping and failure. As a result, no instability or strain localisation can occur, which would lead to the formation of compaction bands. It is believed that cementation plays a key role in controlling strain localisation via compaction bands, though it is unclear how exactly and more research is needed.
\end{enumerate}

%
%

\appendix
\section{Sample porosity after hydrostatic phase} \label{AppendixA}

Volumetric changes cannot be measured during the initial stage of the hydrostatic phase, up to $P_\mathrm{c}$ = $5$~MPa, as no pore-fluid is introduced into the sample yet and hence sample porosity cannot be calculated. After the introduction of pore-fluid, the volumetric changes are measured during the subsequent application of hydrostatic pressure, which is used to calculate sample porosity. The dependence of starting porosity on confining pressure is shown in Figure \ref{Fig:porosity}.

\renewcommand{\thefigure}{A\arabic{figure}}
\setcounter{figure}{0}

%
\begin{figure}[h]
    \centering
    \includegraphics[width=\linewidth]{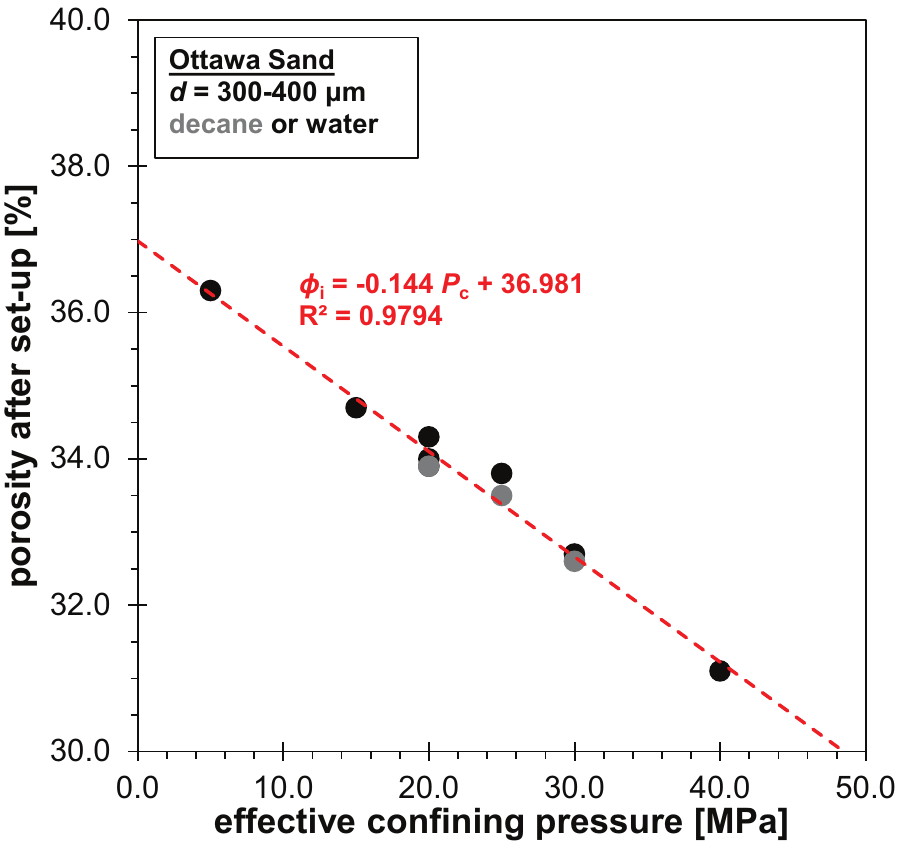}
    \caption{Porosity at the end of the hydrostatic set-up phase as a function of effective confining pressure.}
    \label{Fig:porosity}
\end{figure}

%
%
%
%
%

\begin{acknowledgments}
The authors thank Associate Editor Michele Cooke, Brian Evans and two anonymous Reviewers for their comments on our manuscript. SJTH performed this work as part of the Utrecht University $\textit{Sustainability}$ Research Theme. NB acknowledges support from the UK Natural Environment Research Council, grant number NE/K009656/1. Chris Spiers, Patrick Baud, J\'er\^{o}me Fortin and Antonino Cilona are thanked for their helpful discussions. Experimental data are available from the UK National Geoscience Data Centre (http://www.bgs.ac.uk/services/ngdc/) or upon request to the corresponding author.
\end{acknowledgments}



\end{article}

\end{document}